\documentclass[prd,aps,nofootinbib,twocolumn,showpacs,floatfix,amssymb,amsmath,nofootinbib]{revtex4-1}
\usepackage{slashed}
\usepackage{epsfig}
\usepackage{multirow}

\newcommand{\beqn}{\begin{eqnarray}}
\newcommand{\eeqn}{\end{eqnarray}}
\newcommand{\eq}[1]{(\ref{#1})}

\renewcommand{\P}{{\cal P}}
\newcommand{\T}{{\cal T}}
\newcommand{\PT}{{\cal {PT}}}

\newcommand{\cH}{{\cal H}}

\newcommand{\cO}{{\cal O}}

\newcommand{\diag}{{\mathrm{diag}}}
\newcommand{\sign}{\,{\mathrm{sign}}}

\newcommand{\ket}[1]{{\left|#1\right\rangle}}

\newcommand{\Matrix}[4]{\left( 
\begin{array}{cc}
#1 & #2 \\
#3 & #4
\end{array}
\right)}

\usepackage{color}
\usepackage[dvipsnames]{xcolor}
\definecolor{purple}{rgb}{0.8,0,0.6}
\definecolor{dark-blue}{rgb}{0,0,0.7}
\definecolor{dark-green}{rgb}{0.2,0.8,0.2}

\begin{document}

\title{The Nielsen-Ninomiya theorem, $\PT$-invariant non-Hermiticity\newline and single 8-shaped Dirac cone}

\author{M.~N.~Chernodub }
\affiliation{CNRS, Laboratoire de Math\'ematiques et Physique Th\'eorique UMR 7350, Universit\'e de Tours, 37200 France}
\affiliation{Laboratory of Physics of Living Matter, Far Eastern Federal University, Sukhanova 8, Vladivostok, 690950, Russia}

\begin{abstract}
The Nielsen-Ninomiya theorem implies that any local, Hermitian and translationally invariant lattice action in even-dimensional spacetime possess an equal number of left- and right-handed chiral fermions. We argue that if one sacrifices the property of Hermiticity while keeping the locality and translation invariance, while imposing invariance of the action under the space-time ($\PT$) reversal symmetry, then the excitation spectrum of the theory may contain a non-equal number of left- and right-handed massless fermions with real-valued dispersion. We illustrate our statement in a simple 1+1 dimensional lattice model which exhibits a skewed 8-figure patterns in its energy spectrum. A drawback of the model is that the $\PT$ symmetry of the Hamiltonian is spontaneously broken implying that the energy spectrum contains complex branches. We also demonstrate that the Dirac cone is robust against local disorder so that the massless excitations in this $\PT$ invariant model are not gapped by random space-dependent perturbations in the couplings.
\end{abstract}


\maketitle

\section{Introduction}

Chiral fermions play un increasingly important role in modern physics. One can mention (nearly) massless neutrinos in (astro)particle physics and cosmology~\cite{ref:neutrinos}, light quarks in quark-gluon plasma which emerges in early Universe and heavy ion collisions, condensed matter physics of graphene~\cite{ref:graphene}, Weyl/Dirac semimetals~\cite{ref:semimetals} and liquid helium~\cite{ref:helium}. The massless fermionic excitations may appear in discretized spaces such as naturally ordered crystal structures of real materials in solid state physics or in the lattice models that are used to simulate nonperturbative theories of particle physics. 

One of the most fundamental statements in physics of lattice chiral fermions is the Nielsen-Ninomiya theorem which states that in every physically meaningful discretized theory in even space-time dimensions the chiral fermions should always come in pairs of left- and right-handed chiralities thus keeping the net chirality of the lattice fermions equal to zero~\cite{ref:Nielsen:Ninomiya}. In many cases the fermion doubling is an undesirable property that blocks investigation of certain interesting systems that possess excitations (particles) of only one chirality (for example, neutrinos, which are always left-handed). 

Since the Nielsen-Ninomiya no-go theorem is applied to a broad class of chiral lattice Hamiltonians that are (i) translationally invariant, (ii) local and (iii) Hermitian operators, there were various attempts to circumvent the theorem by abandoning the Hermiticity property of the model while keeping the other requirements. 

The early ideas to avoid the fermion doubling~\cite{Weingarten:1985vc,Alonso:1987ra} with the help of various non-Hermitian Hamiltonians were dismissed either because at a subsequent closer look the models turn out to exhibit doubling~\cite{Gross:1987by,Funakubo:1991sc} or due to emerging inconsistencies at the level of perturbation theory~\cite{Hernandez:1986st}. Moreover, the non-Hermitian Hamiltonians usually have only complex energy spectra thus making their treatment and interpretation of the results rather difficult. 

However, it was later found that a large special class of non-Hermitian Hamiltonians, $H \neq H^\dagger$, unexpectedly possess entirely real energy spectrum~\cite{Bender:1998ke}. These local and translationally invariant Hamiltonians are required to be invariant under a combined action of the space ($\P$) and time ($\T$) reversals, $[\PT,H]=0$. Their energy spectrum stays real apart from certain cases in which the $\PT$ symmetry is broken spontaneously~\cite{Bender:2007nj}. 

A rapid development of experimental technology led to a surge of interest in $\PT$-invariant non-Hermitian systems, both in theoretical and experimental communities, covering broad areas of photonic crystals~\cite{ref:open:photonics1,ref:open:photonics2,ref:open:photonics3,ref:open:photonics4}, non-topological superconductors~\cite{ref:nontop:insulators}, ultracold atoms~\cite{ref:ultracold} to mention a few.

In our paper we attempt to circumvent the Nielsen-Ninomiya theorem by constructing a $\PT$-invariant non-Hermitian Hamiltonian that possesses fermionic excitations of the same handedness. Our choice is supported by the following three arguments. First, the non-Hermitian nature of the Hamiltonian makes it impossible to apply the Nielsen-Ninomiya theorem so that the net number of right- and left-handed fermions may be nonzero. Second, the $\PT$-invariance may guarantee that (at least, a part of) the energy spectrum of these fermions is real. And, finally, third, we notice that many $\PT$ systems naturally possess unidirectional transport~\cite{ref:open:photonics2,ref:open:photonics3,ref:invisibility} what matches perfectly with expected properties of the systems with the fermions of the same handedness.\footnote{For example, in (1+1) dimensions right- and left-handed fer\-mi\-ons are associated with, respectively, right- and left-movers. If the spectrum contains fermions of the same (say, right) handedness then the fermion-mediated transport in the system would be unidirectional (in our example, in the right direction only).} The non-Hermiticity is often associated with unidirectionality~\cite{Chernodub:2015ova}. 

The structure of our paper is as follows. In Sect.~\ref{sec:PT:invariance} we briefly review the $\PT$ symmetry in a continuum theory following Ref.~\cite{Bender:2007nj} and then we discuss its realization in two simplest tight-binding chain Hamiltonians in one spatial dimension. In Sect.~\ref{sec:our:model} we calculate the energy spectrum of a suitable infinite-chain model and show that in certain parameter range the spectrum contains a tilted Dirac cone with lines closed in a figure-8 curve. We show that the fermion solutions corresponding to linear dispersions have the same handedness. In addition to the 8-type Dirac curve the energy spectrum contains two complex arcs which connect upper and lowers parts of the ``8''. Our conclusions are given in the last section.

\section{Hermiticity vs $\P\T$-invariance}
\label{sec:PT:invariance}

\subsection{$\PT$ symmetry in continuum theory}

The parity operator $\P$ flips all spatial coordinates and, consequently, inverts the signs of coordinate and momentum operators,
\beqn
\P {\hat x} \P = - {\hat x}\,, \qquad \P {\hat p} \P = - {\hat p}\,,
\label{eq:P:transformation}
\eeqn
thus leaving intact the commutation relation:
\beqn
[{\hat x},{\hat p}] = i\,.
\label{eq:commutation}
\eeqn

The time-reversal operator $\P$ flips the temporal coordinate, $t \to - t$. It leaves the spatial coordinate unchanged while reversing the sign of the momentum:
\beqn
\T {\hat x} \T = {\hat x}\,, \qquad \T {\hat p} \T = - {\hat p}\,.
\label{eq:T:transformation}
\eeqn
In order to keep compatibility of the $\T$ operation~\eq{eq:T:transformation} with the commutation relation~\eq{eq:commutation}, the time-reversal operation must also change the sign of the complex number $i$:
\beqn
\T i \T = - i\,.
\label{eq:T:i}
\eeqn
In particular, for a complex number $c$ one gets:
\beqn
\T c \T = c^*\,.
\label{eq:T:c}
\eeqn
Relations~\eq{eq:T:i} and \eq{eq:T:c} demonstrate that the time-reversal $\T$ is not, actually, a linear operator (it is often called as ``anti-linear'' operator). 

The operators $\P$ and $\T$ commute with each other, $[\P,\T] = 0$, and, obviously, $\P^2 = \T^{\,2} = 1$. For brevity, a $\P\T$ transformation of an operator $\cO$ is defined as follows:
\beqn
\cO^{\P\T} = (\P\T) \cO (\P\T)\,.
\label{eq:O:PT}
\eeqn
A $\PT$-invariant Hamiltonian obeys $H = H^\PT$.

Below we will examine a difference between the property of Hermiticity and the $\P\T$-invariance in tight-binding lattice systems. For simplicity, we work with one-dimensional lattice Hamiltonians.

\subsection{Chain model with one type of sites}

\begin{figure}[!thb]
\includegraphics[scale=0.45,clip=true]{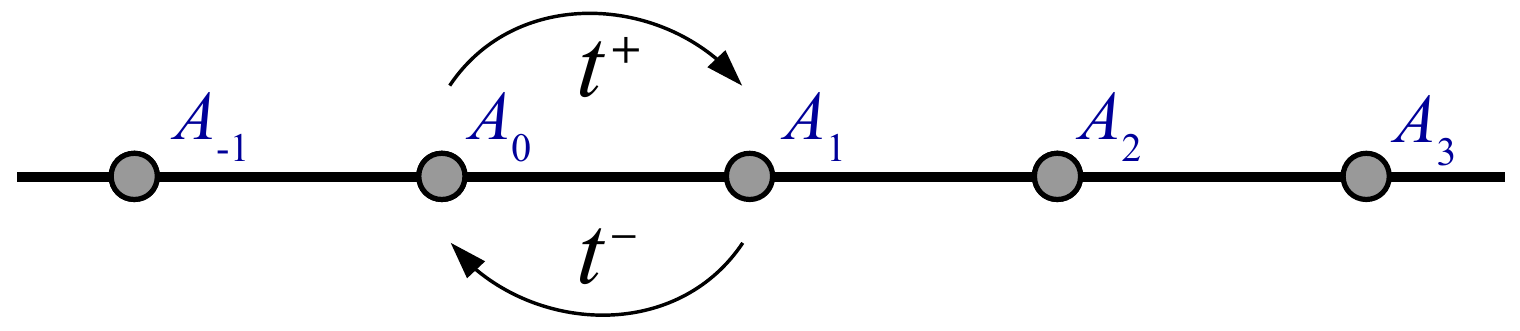}
\caption{Uniform one-dimensional chain with one type of sites described by the Hamiltonian~\eq{eq:H:chain1}.}
\label{fig:chain1}
\end{figure}

Consider first a simplest chain Hamiltonian,
\beqn
H_1 = - \sum_l \left( t^+ a_{l+1}^\dagger a_l + t^- a_{l-1}^\dagger a_l \right)\,,
\label{eq:H:chain1}
\eeqn
with one type of lattice sites $A_l$ labelled by the index~$l$, Fig.~\ref{fig:chain1}. Here $a^\dagger_l$ and $a_l$ are, respectively, creation and annihilation operators satisfying the fermionic anticommutation relation $\{a_{l}, a^\dagger_{l'}\} = \delta_{ll'}$. The hopping parameters $t^+$ and $t^-$ correspond to the hops of electrons in positive and negative directions along the chain lattice, Fig.~\ref{fig:chain1}. 

Applying a Hermitian conjugation to the Hamiltonian~\eq{eq:H:chain1} one gets:
\beqn
H_1^\dagger = - \sum_l \left[ (t^+)^* \, a_{l-1}^\dagger a_l + (t^-)^*\, a_{l+1}^\dagger a_l \right]\,,
\label{eq:H:chain1:h}
\eeqn
where we rearranged the variable $l \to l \pm 1$ appropriately.

According to Eqs.~\eq{eq:H:chain1} and \eq{eq:H:chain1:h} the Hermiticity of the Hamiltonian, $H_1^\dagger = H_1$, implies the following simple relation between the hopping parameters:
\beqn
\mbox{Hermiticity:}\qquad t^- = (t^+)^*\,.
\label{eq:tt:Hermiticity}
\eeqn

The parity transformation~\eq{eq:P:transformation} and the time reversal~\eq{eq:T:transformation} and \eq{eq:T:i} act on creation/annihilation operators and the hopping parameter as follows:
\beqn
\P a_l \P & = & a_{-l}\,, 
\qquad 
\P a^\dagger_l \P = a^\dagger_{-l}\,,  
\qquad
\P t^\pm \P = t^\pm\,,
\label{eq:P:one:chain}
\\
\T a_l \T & = & a_l\,, 
\qquad 
\T a^\dagger_l \T = a^\dagger_l\,,  
\qquad
\T t^\pm \T = (t^\pm)^*\,.
\quad
\label{eq:P:two:chain}
\eeqn
The $\P\T$ transformed Hamiltonian~\eq{eq:H:chain1} then reads,
\beqn
H_1^{\P\T} = - \sum_l \left[ (t^+)^* a_{l-1}^\dagger a_l + (t^-)^* a_{l+1}^\dagger a_l \right]\,,
\label{eq:H:chain1:PT}
\eeqn
where we have inverted the summation variable $l \to - l$.

Comparing the Hamiltonians~\eq{eq:H:chain1:PT} and \eq{eq:H:chain1:h} we conclude that for the simplest chain model~\eq{eq:H:chain1:h} the $\P\T$ invariance imposes the following condition on the hopping parameters:
\beqn
\mbox{$\P\T$ invariance:}\qquad t^- = (t^+)^*\,.
\label{eq:tt:PT}
\eeqn
which coincides with the condition for Hermiticity~\eq{eq:tt:Hermiticity}. Therefore, in the simplest uniform model~\eq{eq:H:chain1} the $\P\T$ invariance implies Hermiticity and vice versa, $H_1^{\P\T} \equiv H_1^\dagger$. In other words, it is impossible to construct simultaneously non-Hermitian and  $\P\T$ invariant model using only one (sub)lattice of sites. 

\subsection{Chain model with two sublattices}

\subsubsection{The model}

\begin{figure}[!thb]
\includegraphics[scale=0.35,clip=true]{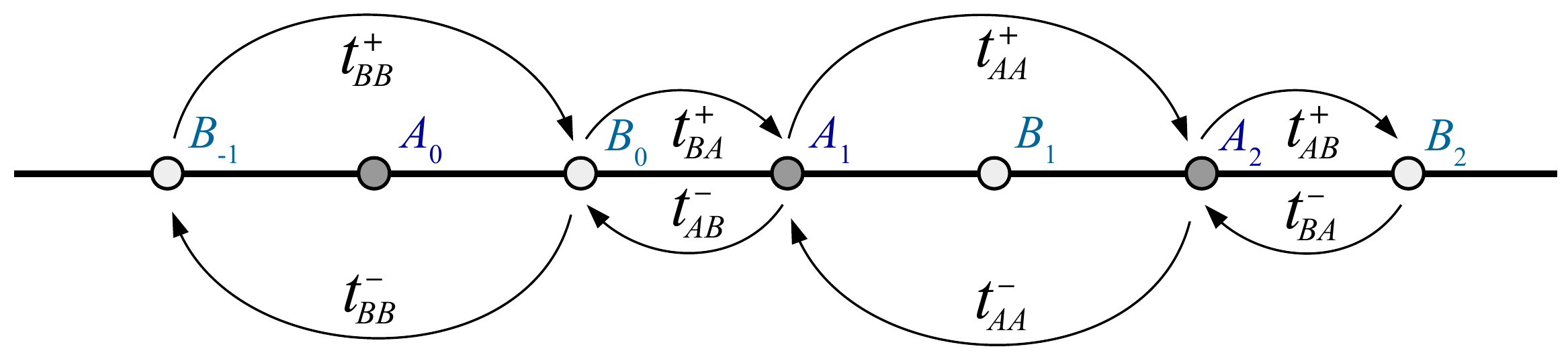}
\caption{Uniform one-dimensional chain with two types of sites described by the Hamiltonian~\eq{eq:H:chain2}.}
\label{fig:chain2}
\end{figure}

Now let us consider a model with sublattices of alternating lattice sites, $A$ and $B$, as shown in Fig.~\ref{fig:chain2}. 

We allow for the electron to hop between the nearest-neighbor sites, both in positive and negative directions, from an $A$ site to a nearest $B$ site, and vise versa. The corresponding hoping parameters are $t^\pm_{AB}$ and $t^\pm_{BA}$, respectively. Here the superscript ($\pm$) marks the direction of the hop and the superscript ($AB$ or $BA$) corresponds to the starting and ending sites of the hop. We also allow to an electron to make hops between sites of the same type, by adding into consideration the next-to-the-nearest  hopping parameters $t^\pm_{AA}$ and $t^\pm_{BB}$, which describe nearest-neighbor motion within the same sublattice.

The corresponding Hamiltonian is:
\beqn
H_2 = - \sum_l \Bigl[ & & t_{AB}^+ b^\dagger_l a_l + t_{AB}^- b^\dagger_{l-1} a_l \nonumber \\[-2mm]
+ & & t_{BA}^+ a^\dagger_{l+1} b_l + t_{BA}^- a^\dagger_l b_l \nonumber \\
+ & &  t^+_{AA} a_{l+1}^\dagger a_l + t^-_{AA} a_{l-1}^\dagger a_l  \nonumber \\
+ & &  t^+_{BB} b_{l+1}^\dagger b_l + t^-_{BB} b_{l-1}^\dagger b_l  \Bigr]\,,
\label{eq:H:chain2}
\eeqn
where $a_l$ ($a^\dagger_l$) and $b_l$ ($b^\dagger_l$) are the annihilation and creation operators at the $A$ and $B$ sublattices, respectively. 

The four lines in the Hamiltonian~\eq{eq:H:chain2:h} describe the nearest-neighbor hops, respectively, (i) from the sublattice $A$ to $B$, (ii) from the sublattice $B$ to $A$; (iii) within the $A$ sublattice and (iv) within the $B$ sublattice. First (second) terms in each line corresponds to hops in positive (negative) direction.

\subsubsection{Hermiticity}

Applying the Hermitian conjugation to the Hamiltonian~\eq{eq:H:chain2} one gets:
\beqn
H_2^\dagger = - \sum_l \Bigl[ & &  (t_{AB}^+)^*\, a^\dagger_l b_l + (t_{AB}^-)^*\, a^\dagger_l b_{l-1}  \nonumber \\[-2mm]
+ & &  (t_{BA}^+)^*\,b^\dagger_l a_{l+1} + (t_{BA}^-)^*\, b^\dagger_l a_l \nonumber \\
+ & &  (t^+_{AA})^*\,a^\dagger_l a_{l+1} + (t^-_{AA})^*\, a^\dagger_l a_{l-1}  \nonumber \\
+ & &  (t^+_{BB})^*\,b^\dagger_l b_{l+1} + (t^-_{BB})^*\, b^\dagger_l b_{l-1}  \Bigr]\,,
\label{eq:H:chain2:h}
\eeqn
where we rearranged the variable $l \to l \pm 1$ appropriately.

According to Eqs.~\eq{eq:H:chain2} and \eq{eq:H:chain2:h} the requirement of Hermiticity of the Hamiltonian~\eq{eq:H:chain2}, $H_2^\dagger = H_2$, implies the following set of relations between the hopping parameters:
\beqn
\mbox{Hermiticity:}\qquad \left\{
\begin{array}{rcl}
\left( t_{AB}^+ \right)^* & = & t_{BA}^- \\
\left( t_{BA}^+ \right)^* & = & t_{AB}^- \\
\left( t_{AA}^+ \right)^* & = & t_{AA}^- \\
\left( t_{BB}^+ \right)^* & = & t_{BB}^-
\end{array}
\right.\,.
\label{eq:tt:Hermiticity:chain2}
\eeqn
The hopping parameters corresponding to the hops forward and backwards are related by the complex conjugation similarly to the simplest uniform model~\eq{eq:tt:PT}.

\subsubsection{$\P\T$ invariance}

\begin{figure}[!thb]
\includegraphics[scale=0.3,clip=true]{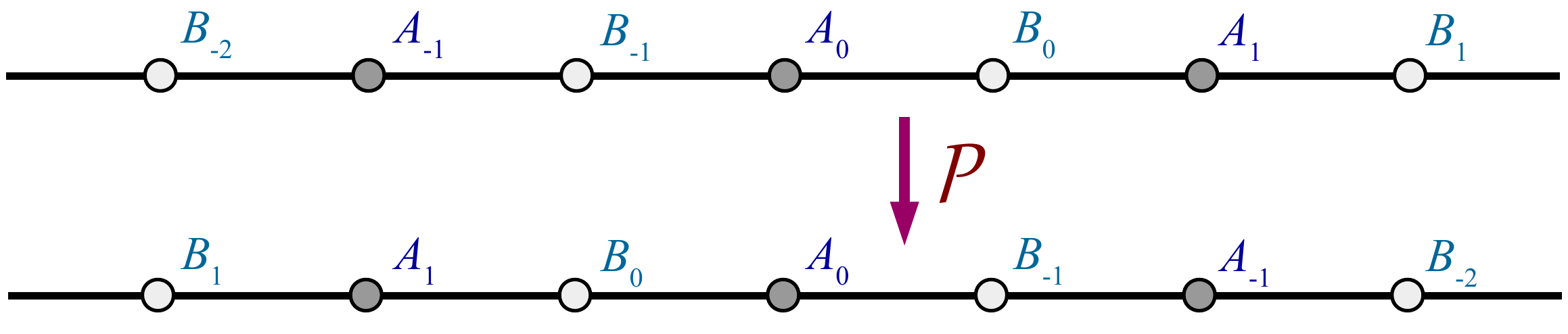}
\caption{The parity transformation~$\P$, Eq.~\eq{eq:P:transformation}, applied to the alternating lattice of Fig.~\ref{fig:chain2}.}
\label{fig:chain2:inversion}
\end{figure}

The $\P$ parity inversion~\eq{eq:P:transformation} works slightly differently for the $A$ and $B$ sublattices. From Fig.~\ref{fig:chain2:inversion} we deduce that the parity $\P$ flip~\eq{eq:P:transformation} acts on the creation/annihilation operators as follows:
\beqn
& & \P a_l \P = a_{-l}\,, \qquad  \ \ \, \P a^\dagger_l \P = a^\dagger_{-l}\,, \\
& & \P b_l \P = b_{-l-1}\,, \qquad  \P b^\dagger_l \P = b^\dagger_{-l-1}\,,
\eeqn
while the time-reversal operation $\T$, Eq.~\eq{eq:T:transformation}, leaves them intact:
\beqn
& & \T a_l \T = a_{l}\,, \qquad  \T a^\dagger_l \T = a^\dagger_{l}\,, \\
& & \T b_l \T = b_{l}\,, \qquad\,  \T b^\dagger_l \T = b^\dagger_{l}\,.
\eeqn

Similarly to the case of the uniform Hamiltonian $H_1$, the complex hopping parameters are not affected by the parity transformation while they are changed to their complex conjugated after the time-reversal operation~\eq{eq:T:c} is applied:
\beqn
 \P t^\pm_{\ell\ell'} \P = t^\pm_{\ell\ell'}, \qquad \T t^\pm_{\ell\ell'} \T = \left(t^\pm_{\ell\ell'}\right)^*, \quad \ell\ell' = A, B. \quad
\label{eq:t:PT:chain2}
\eeqn

The $\P\T$ transformed Hamiltonian~\eq{eq:H:chain2} takes the form:
\beqn
H_2^{\P\T} = - \sum_l \Bigl[ & & (t_{AB}^+)^*\,  b^\dagger_{-l-1} a_{-l} + (t_{AB}^-)^*\,   b^\dagger_{-l} a_{-l} \label{eq:H:chain2:PV:1}
 \\[-2mm]
+ & & (t_{BA}^+)^*\,   a^\dagger_{-l-1} b_{-l-1} + (t_{BA}^-)^*\,   a^\dagger_{-l} b_{-l-1} \nonumber \\
+ & &  (t^+_{AA})^*\,   a_{-l-1}^\dagger a_{-l} + (t^-_{AA})^*\,   a_{-l+1}^\dagger a_{-l}  \nonumber \\
+ & &  (t^+_{BB})^*\,   b_{-l-2}^\dagger b_{-l-1} + (t^-_{BB})^*\,   b_{-l}^\dagger b_{-l-1}  \Bigr]. \nonumber
\eeqn
Changing the summation variable $l \to - l$ or $l \to - l +1$ for certain lines in Eq.~\eq{eq:H:chain2:PV:1}, we get:
\beqn
H_2^{\P\T} = - \sum_l \Bigl[ & & (t_{AB}^+)^*\,  b^\dagger_{l-1} a_{l} + (t_{AB}^-)^*\,   b^\dagger_{l} a_{l} 
\nonumber\\[-2mm]
+ & & (t_{BA}^+)^*\,   a^\dagger_{l} b_{l} + (t_{BA}^-)^*\,   a^\dagger_{l+1} b_{l} \nonumber \\
+ & &  (t^+_{AA})^*\,   a_{l-1}^\dagger a_{l} + (t^-_{AA})^*\,   a_{l+1}^\dagger a_{l}  \nonumber \\
+ & &  (t^+_{BB})^*\,   b_{l-1}^\dagger b_{l} + (t^-_{BB})^*\,   b_{l+1}^\dagger b_{l}  \Bigr]. 
\label{eq:H:chain2:PV}
\eeqn

Next, we compare the Hamiltonians~\eq{eq:H:chain2} and \eq{eq:H:chain2:PV}, and conclude that the $\P\T$ invariance $H^{\P\T}_2 = H_2$ imposes the following condition on the hopping parameters:
\beqn
\mbox{$\P\T$ invariance:}\qquad \left\{
\begin{array}{rcl}
\left( t_{AB}^+ \right)^* & = & t_{AB}^- \\
\left( t_{BA}^+ \right)^* & = & t_{BA}^- \\
\left( t_{AA}^+ \right)^* & = & t_{AA}^- \\
\left( t_{BB}^+ \right)^* & = & t_{BB}^-
\end{array}
\right.\,.
\label{eq:tt:PT:chain2}
\eeqn

The constrains that are imposed on the hopping parameters by the Hermiticity~\eq{eq:tt:Hermiticity:chain2} and by the $\P\T$ invariance \eq{eq:tt:PT:chain2} are not equivalent. In more details, the intra-lattice hopping parameters between the sites of the same sublattices ($t^\pm_{AA}$ and $t^\pm_{BB}$) are, unsurprisingly, obeying the same conditions both for Hermitian and $\P\T$ symmetric Hamiltonians: the hops forward and backward are related to each other by the complex conjugation. However, the parameters that control inter-lattice hopping ($t^\pm_{AB}$ and $t^\pm_{BA}$) satisfy different relations for Hermitian and $\P\T$ invariant Hamiltonians. Therefore the model with two sublattices~\eq{eq:H:chain2} allows us to construct a chain Hamiltonian which is not Hermitian while being $\PT$ invariant at the same time.

\section{The model}
\label{sec:our:model}

\subsection{$\PT$ invariant non-Hermitian Hamiltonian}

We are interested in a $\P\T$ symmetric non-Hermitian Hamiltonian. We have eight complex parameters $t^\pm_{\ell\ell'}$ with $\ell,\ell' = A,B$, which are related to each other by four equations~\eq{eq:tt:PT:chain2}. Thus we have four independent hopping parameters, or eight real parameters that describe our model. For the sake of convenience we reduce the parameter space and choose the following set of parameters:
\beqn
& & t^+_{AA} = - t^-_{AA} = i t_A\,,  \qquad t^+_{BB} = - t^-_{BB} = i t_B\,,  
\label{eq:intralattice}\\
& & t^+_{AB} = - t^-_{AB} = i g_1\,,  \qquad t^+_{BA} = - t^-_{BA} = i g_2\,,
\label{eq:interlattice}
\eeqn
where $t_A$, $t_B$, $g_1$ and $g_2$ are real-valued numbers.

The Hamiltonian~\eq{eq:H:chain2} with the hopping parameters~\eq{eq:intralattice} and \eq{eq:interlattice} reads as follows:
\beqn
H & = & - i \sum_l \Bigl[ g_1 \bigl(  b^\dagger_l a_l - b^\dagger_{l-1} a_l \bigr) + g_2 \bigl( a^\dagger_{l+1} b_l - a^\dagger_l b_l\bigr) \nonumber \\
&&  + t_A (a_{l+1}^\dagger a_l - a_{l-1}^\dagger a_l) + t_B (b_{l+1}^\dagger b_l - b_{l-1}^\dagger b_l) \Bigr]. \quad \quad
\label{eq:H:chain2:explicit}
\eeqn

By construction, the Hamiltonian~\eq{eq:H:chain2:explicit} is a $\PT$ invariant but not Hermitian operator. The special choice of parameters~\eq{eq:intralattice} and \eq{eq:interlattice} allows us to determine the Hermiticity criteria in the simple way: the intra-lattice coupling parameters~\eq{eq:intralattice} of the $\P\T$-invariant Hamiltonian satisfy the Hermiticity conditions~\eq{eq:tt:Hermiticity:chain2} automatically while the inter-lattice hopping parameters~\eq{eq:interlattice} satisfy the Hermiticity conditions~\eq{eq:tt:Hermiticity:chain2} if and only if $g_1 = g_2$:
\beqn
\mbox{Hamiltonian:} \quad \left\{ 
\begin{array}{rcl}
g_1 & = & g_2\,,\qquad \mbox{Hermitian}\,, \\[1mm]
g_1 & \neq & g_2\,,\qquad \mbox{non-Hermitian}\,.
\end{array}
\right.\quad
\label{eq:Hermiticity:g1g2}
\eeqn

\subsection{Energy spectrum}

We look for eigenstates of the Hamiltonian~\eq{eq:H:chain2:explicit} in the standard form:
\beqn
\ket{\psi} = \sum_{l} \left( \alpha_l a^\dagger_l + \beta_l b^\dagger_l  \right) \ket{0}\,,
\label{eq:trial}
\eeqn
where $\ket{0}$ is the vacuum state with $a_l \ket{0} = b_l \ket{0} = 0$.

Applying Eq.~\eq{eq:H:chain2:explicit} to \eq{eq:trial} we solve the eigenstate equation $H \ket{\psi} = \epsilon \ket{\psi}$ using the ansatz 
\beqn
\alpha_l = \alpha e^{i p l}\,, 
\qquad
\beta_l = \beta e^{i p l}\,,
\eeqn
where $p$ is a quasimomentum and $\alpha$ and $\beta$ are complex numbers. The eigensystem is determined by the following matrix equation:
\beqn
\cH_p \Psi_p = \epsilon(p) \Psi_p\,,
\label{eq:H:p:eq}
\eeqn
where
\beqn
\cH_p = 
- 2 \left( 
\begin{array}{ll}
t_A \sin p & \qquad g_2 \, e^{- \frac{i p}{2}} \sin\frac{p}{2}\\
g_1 \, e^{\frac{i p}{2}} \sin\frac{p}{2} & \qquad t_B \sin p \\
\end{array}
\right)\,, 
\label{eq:Hp}
\eeqn
and 
\beqn
\Psi_p = 
\left( 
\begin{array}{l}
\alpha\\
\beta
\end{array}
\right)\,.
\label{eq:Phi}
\eeqn
In accordance with Eq.~\eq{eq:Hermiticity:g1g2} the Hamiltonian $\cH_p$ in Eq.~\eq{eq:Hp} is indeed a Hermitian operator, $\cH_p^\dagger = \cH_p$, provided $g_1 = g_2$.

The eigenenergies $\epsilon_{\pm}(p)$ can be readily determined from Eq.~\eq{eq:H:p:eq}:
\beqn
& & \epsilon_{\pm}(p) = - ( t_A + t_B ) \sin p \nonumber \\
& & \ \ \qquad \pm \sign(p) \sqrt{ ( t_A - t_B )^2 \sin^2 p + 4 h \sin^2 \frac{p}{2}}\,,
\label{eq:E:pm}
\eeqn
where 
\beqn
h = g_1 g_2\,.
\label{eq:h}
\eeqn
For Hermitian Hamiltonians one always has $h \geqslant 0$ while the region $h<0$ corresponds to a non-Hermitian case.

\subsection{Weyl modes in a single closed Dirac cone}
\label{sec:Dirac:modes}

For a generic set of parameters ($t_A,t_B,h$) our model \eq{eq:H:chain2:explicit} possesses gapless solutions  $\epsilon_{\pm}(p) = \epsilon^{(0)}_{\pm}(p) + O(p^3)$ with a linear dispersion relation at the origin $p=0$:
\beqn
\epsilon^{(0)}_{\pm}(p) = v_\pm \, p\,.
\label{eq:E0:pm}
\eeqn
According to Eq.~\eq{eq:E:pm} these modes are propagating with the following velocities:
\beqn
v_{\pm} = - (t_A + t_B) \pm \sqrt{(t_A - t_B)^2 + h}\,.
\label{eq:vpm}
\eeqn

\begin{figure*}[!thb]
\begin{center}
\includegraphics[scale=0.5,clip=true]{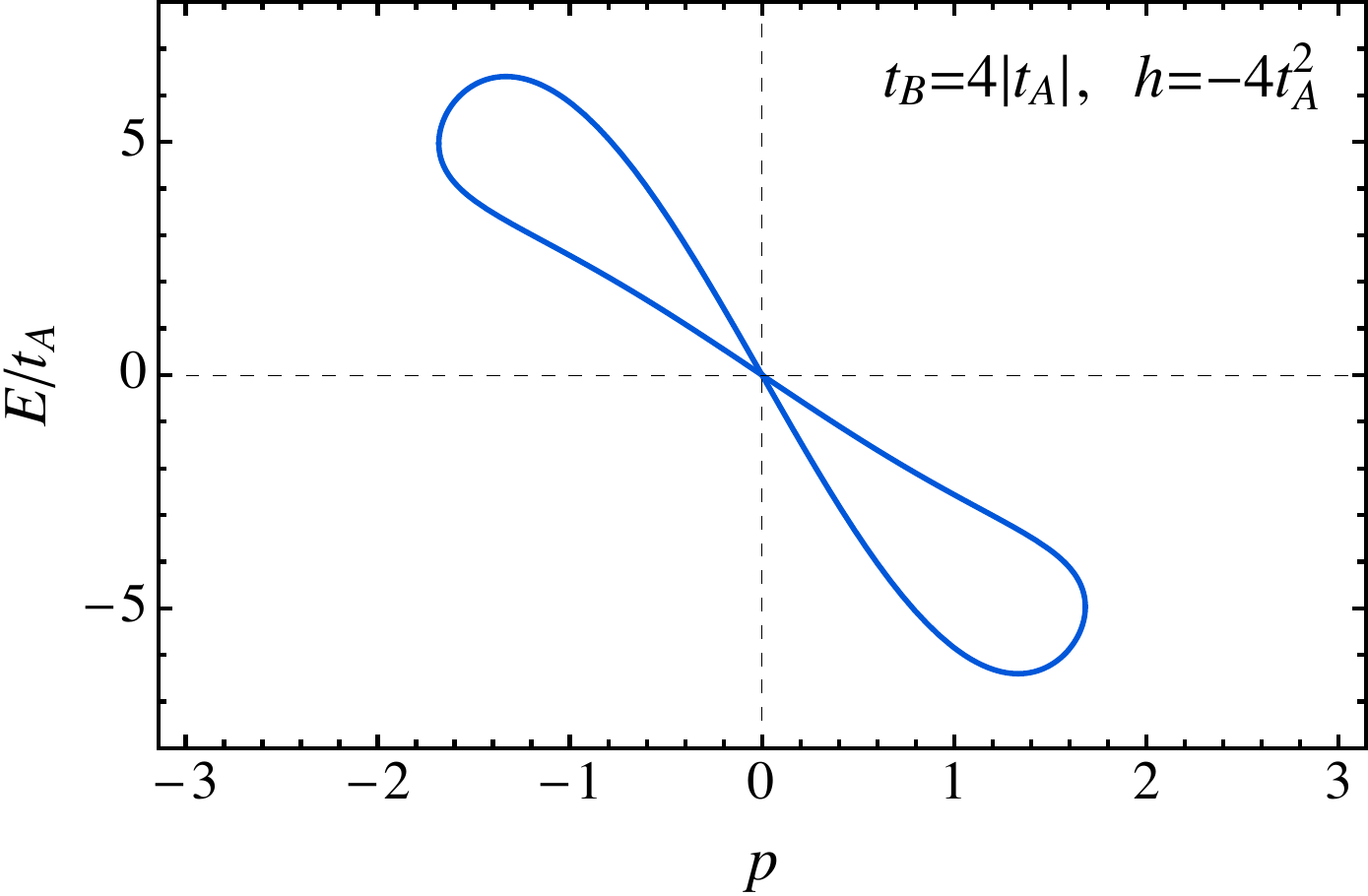}
\hskip 17mm
\includegraphics[scale=0.45,clip=true]{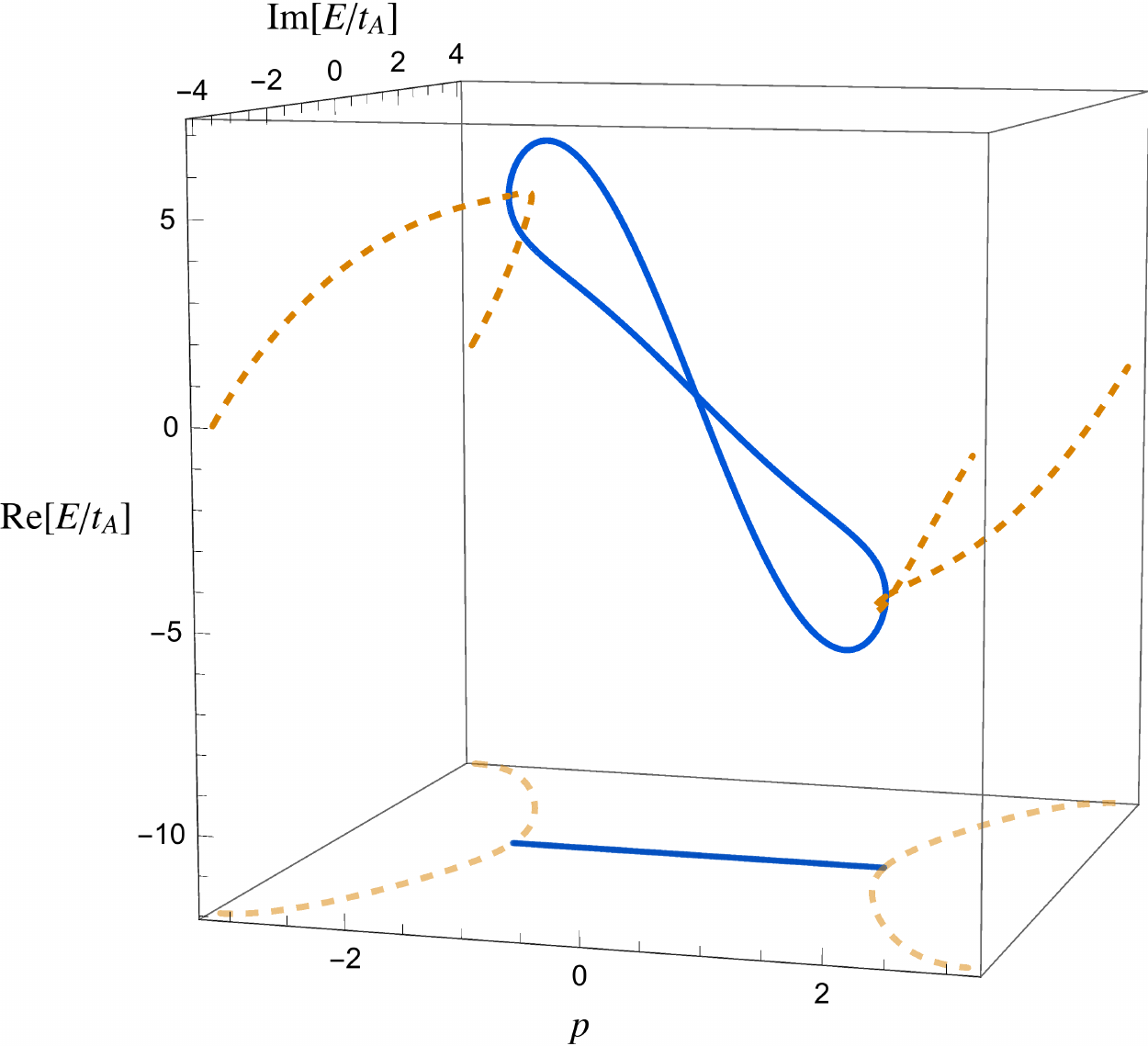}
\end{center}
\caption{The energy spectrum~\eq{eq:E:pm} with $t_A = 1$, $t_B = 1/4$ and $h = - 1/4$. The real-valued branches of energy form the 8-type closed curve (shown by the blue line) while the complex valued energy has a topology of the circle (shown by the orange line).}
\label{fig:spectrum:example}
\end{figure*}

However, what is most interesting is the global behavior of the dispersions~\eq{eq:E:pm} in the Brillouin zone. For a certain region of the parameters the energy spectrum~\eq{eq:E:pm} has a real-valued dispersion relation which has a shape of a skewed 8-figure as it is shown by the blue solid line in an example of Fig.~\ref{fig:spectrum:example}. The upper and lower parts of the 8-shaped energy curve are attached to each other by two a complex-valued energy arks (shown by the dashed orange lines in Fig.~\ref{fig:spectrum:example}). 

Intriguingly, the 8-shaped spectrum describes massless fermions, and there is precisely {\emph {one}} Dirac cone in the whole Brillouin zone with the real-valued energy dispersion. This fact means that if massless excitations at the cone have the same handedness then they cannot be compensated by excitations from another cone with a different handedness. Below we explicitly demonstrate that the gapless modes~\eq{eq:E0:pm} are indeed fermions which can be described by a $1+1$ dimensional Dirac equation, and that these modes have indeed the same handedness in a certain range of the model parameters.

We expand the Hamiltonian around the origin, $\cH_p = \cH^{(0)}_p + O(p^3)$, and get the linearized Hamiltonian 
\beqn
\cH^{(0)}_p = -
\left( 
\begin{array}{ll}
2 t_A &  g_2\\
g_1  & 2 t_B \\
\end{array}
\right) p \,.
\label{eq:Hp:low:p}
\eeqn
that acts on two-component wavefunction~\eq{eq:Phi}:
\beqn
\cH^{(0)}_p \Psi^{(0)}_p = \epsilon_p \Psi^{(0)}_p\,.
\label{eq:cH:0}
\eeqn
The corresponding linear dispersions $\epsilon_p \equiv \epsilon^{(0)}_{\pm}$ given in Eqs.~\eq{eq:E0:pm} and \eq{eq:vpm}. The superscript ``$(0)$'' indicates that Eq.~\eq{eq:cH:0} is valid in the vicinity of the origin $p=0$.

Equation~\eq{eq:cH:0} can be interpreted in terms of a Dirac equation in which the fermion spin, similarly to graphene, corresponds to an internal degree of freedom associated with the presence of two sublattices $A$ and $B$. We choose the following representation for the two-dimensional Dirac matrices:
\beqn
\gamma^0 
= \Matrix{0}{1}{1}{0}\,,
\qquad
\gamma^1 
=  \Matrix{0}{1}{-1}{0}\,.
\label{eq:gamma:matrix}
\eeqn
The matrices~\eq{eq:gamma:matrix} satisfy the commutation relations $\{\gamma^\mu,\gamma^\nu\} = 2 g^{\mu\nu}$, where $g^{\mu\nu} = \diag(1,-1)$ is the metric tensor. The analogue of the $\gamma^5$ matrix in (1+1) dimensions is the following~\cite{Harvey:2005it}:
\beqn
{\bar \gamma} = \gamma^0 \gamma^1 = \sigma_z \equiv \Matrix{1}{0}{0}{-1}\,.
\eeqn

The projectors on positive ($P_+$) and negative ($P_+$) chiralities are:
\beqn
P_\pm = \frac{1\pm {\bar \gamma}}{2}\,,
\eeqn
or, explicitly,
\beqn
P_+ = \Matrix{1}{0}{0}{0}\,, 
\qquad 
P_- = \Matrix{0}{0}{0}{1}\,.
\eeqn
The projectors satisfy the relations $P_+ P_- = P_- P_+ = 0$ and $P_\pm^2 = P_\pm$.

We diagonalize the Hamiltonian~\eq{eq:Hp:low:p},
\beqn
\cH^{(0)}_p = U^\dagger \cH^{\mathrm{diag}}_p U\,,
\label{eq:diagonalizing:H}
\eeqn
where $U$ is an $SU(2)$ rotation matrix and 
\beqn
\cH^{\mathrm{diag}}_p =
\left( 
\begin{array}{ll}
v_+ & 0 \\
0  & v_- \\
\end{array}
\right) p \,,
\label{eq:Hp:low:p}
\eeqn
is the diagonalized Hamiltonian with the velocities~\eq{eq:vpm}. Then we substitute Eq.~\eq{eq:diagonalizing:H} into \eq{eq:cH:0}, and multiply the result by the matrix $\gamma^0 U/(v_+ v_-)$ where $v_+ v_- = 4 t_A t_B- h$. We get for the eigenvalue equation~\eq{eq:cH:0} the following expression:
\beqn
\left[\gamma^\mu \partial_\mu^+ P_+ - \gamma^\mu \partial_\mu^- P_- \right] \psi_p(t,x) = 0 \,,
\eeqn
where the wave function
\beqn
\psi_p(t,x) = \chi_p e^{ - i \epsilon t + i p x}\,,
\eeqn
is expressed via the spinor $\chi_p = U \Psi^{(0)}_p$, and the derivatives are defined as follows:
\beqn
\partial_\mu^\pm = \left( \frac{1}{v_\pm} \frac{\partial}{\partial t}, \frac{\partial}{\partial x} \right)\,.
\eeqn
The solutions with positive and negative chiralities,
\beqn
\psi_\pm = P_\pm \psi\,,
\eeqn
satisfy the following two-component Weyl equations,
\beqn
\gamma^\mu \partial_\mu^\pm \psi_\pm (t,x) = 0\,,
\label{eq:Dirac:eq}
\eeqn
or, in the explicit form:
\beqn 
\left(\gamma^0 \frac{1}{v_\pm} \frac{\partial}{\partial t} + \gamma^1 \frac{\partial}{\partial x} \right) \psi_\pm (t,x) = 0\,.
\label{eq:Dirac:explicit}
\eeqn
Thus, in each Dirac cone we have two Weyl fermions with the dispersions~\eq{eq:E0:pm}, $\epsilon_\pm = v_\pm p$ corresponding to the propagation with velocities~\eq{eq:vpm}.

\subsection{Handedness and chirality of the (1+1) modes}

Before proceeding further it is important to clarify the issue of handedness of fermions in (1+1) dimensions. The fermions have right- of left-handed helicity provided the projection of the fermion's spin onto direction of fermion's  momentum is a positive or negative quantity, respectively. Strictly speaking, in (1+1) dimensions the fermions do not carry a dynamical spin degree of freedom. For example, a fermionic field theory can be bosonized and, consequently, represented by a field theory of spinless bosons. 

In the absence of spin we cannot, in a strict sense, define the helicity in (1+1) dimensions. However, it is customary to associate right-handed and left-handed particles with particles moving in positive (right-movers, $v>0$) and negative (left-movers, $v<0$) directions. To justify this choice, one can imagine, for example, a theory of chiral fermions in $3+1$ dimensions subjected to sufficiently strong magnetic field directed along the $x$ axis. In these conditions 
\begin{itemize}
\item[(i)] the spin of the fermions is pointed along the positive direction of the $z$ axis;
\item[(ii)] the fermions occupy the lowest Landau level; 
\item[(iii)] the motion of the fermions is restricted along the same axis. 
\end{itemize}
As a consequence, right- and left-handed chiral fermions in (3+1) dimensions become, respectively, right- and left-movers in the reduced (1+1) dimensions in the $(t, z)$ plane with the (1+1) dimensional energy dispersion relation $\varepsilon = |p_z|$.

The Dirac equation~\eq{eq:Dirac:explicit} implies that the states with positive and negative chiralities propagate with the velocities $v_+$ and $v_-$, respectively, $x = v_\pm t$. Their dispersion relations are given by Eq.~\eq{eq:E0:pm}. If the velocities $v_+$ and $v_-$ have the same sign, then the handedness of the corresponding branches of solutions is the same because they propagate in the same direction. Therefore, the positive and negative solutions of the chiral $\gamma_5$ operator may be {\emph{both}} right-handed or left-handed from the point of view of their helicity.

\subsection{The parameter space of the model}

The parameter space of the model contains five different regions classified by qualitative features of the energy spectrum. In Fig.~\ref{fig:diagram} we show the energy spectra in different regions of the parameter space of the model~\eq{eq:H:chain2:explicit} labelled by the couplings $t_B$ and $h \equiv g_1 g_2$ (in units of coupling $t_A$). The essential features of the energy spectra are as follows:

\begin{figure*}[!thb]
\begin{center}
\includegraphics[scale=0.6,clip=true]{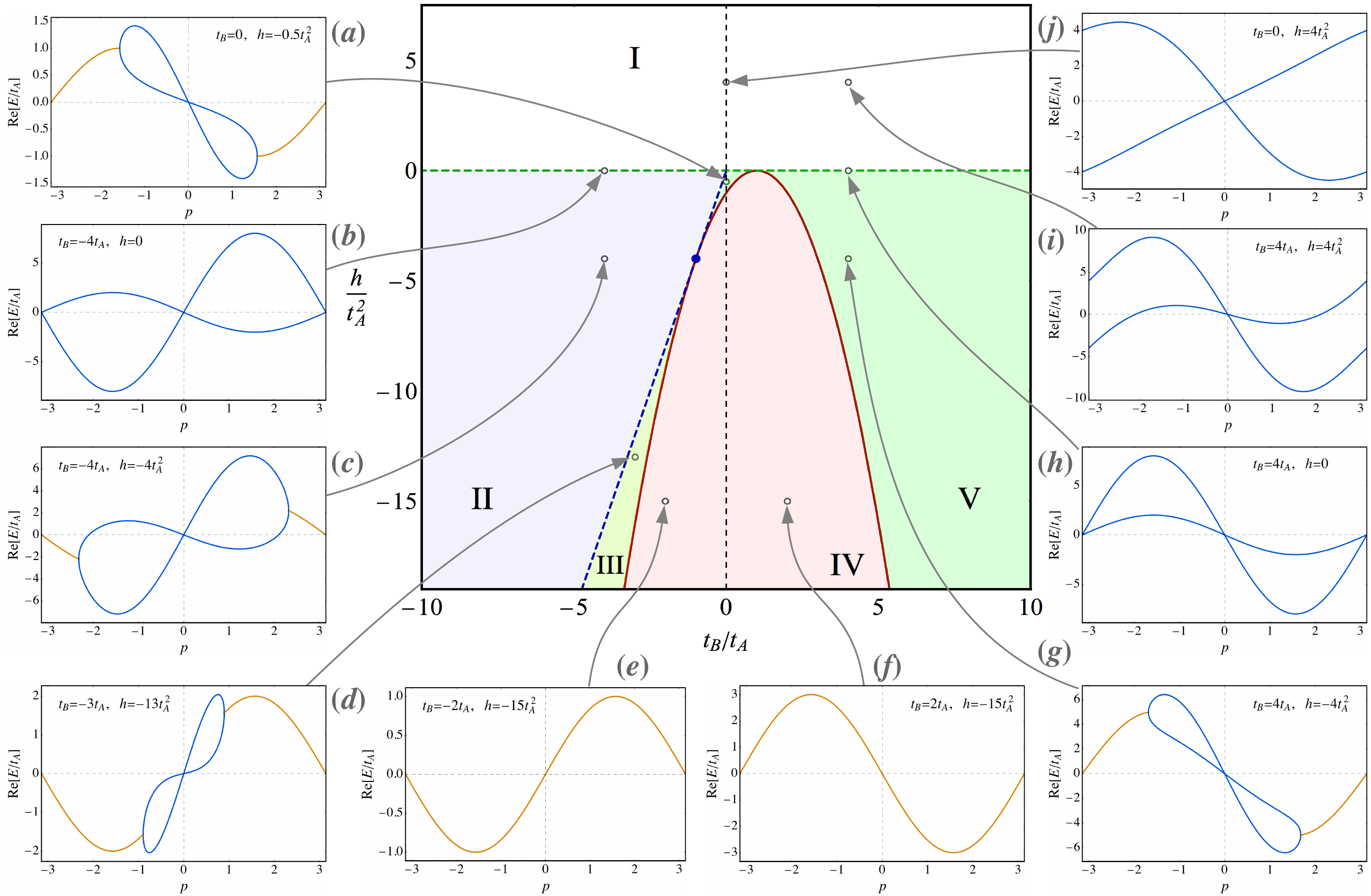}
\end{center}
\caption{Energy spectrum in the space of couplings $t_A$ and $h$ (in units of $t_A > 0$). The model is Hermitian at $h\geqslant0$ and non-Hermitian at $h <0$, the boundary $h=0$ between these semiplanes is marked by horizontal dashed green line. The skewed 8-type spectrum (with fermions of the same helicity) appears only in the regions III and V shaded by the greenish colors. The Dirac cone hosts either two left-handed fermions [examples (a) and (g) in the green area V] or two right-handed fermions [example (d) in the yellow-green area III]. In the red-shaded region IV the spectrum has no real-valued eigenenergies [(e) and (f)], while the blue shaded region II possesses excitations with the 8-type spectrum where a Dirac cone hosts fermions of both helicities (c). In the Hermitian part I the energy spectrum has only real eigenvalues [(b), (h), (i) and (j)]. The regions II and III are separated by the dashed blue line $h = 4 t_A t_B$, while the boundary of the region IV is determined by the relation $h = - (t_A - t_B)^2$ (the red solid curve).}
\label{fig:diagram}
\end{figure*}

\begin{itemize}
\item[$\blacktriangleright$] The model is Hermitian at $h \geqslant 0$ and non-Hermitian at $h<0$.
\item[$\blacktriangleright$] In the upper, Hermitian semiplane (the unshaded \underline{region I} in Fig.~\ref{fig:diagram}) the energy dispersions are given by purely real-valued functions. The typical examples of the energy spectra across this region are shown in the insets (b), (i), (j) and (h).
\item[$\blacktriangleright$] The lower (non-Hermitian) semiplane contains four distinct regions:


\item[$\triangleright$] In \underline{region II} (shaded by the blue color) the Dirac cone at $p=0$ hosts both left-handed and right-handed fermions. The spectrum possesses two complex-valued arcs that connect the parts of the Dirac cone over the boundary at $p = \pm \pi$. An example is shown in plot (c). Region II is determined by the couplings satisfying $0> h > 4 t_A t_B$ and $h<0$.
\item[$\triangleright$] In \underline{region IV} (shaded by the red color) the spectrum does not possess purely real-valued eigenvalues at all. The Dirac cones are located in the imaginary space as seen from plots (e) and (f). Region IV is determined by the couplings $h < - (t_A - t_B)^2$.
\item[$\triangleright$] In \underline{region III} (shaded by the yellow-green color) the clockwise skewed 8-type closed Dirac cone possesses two fermions with the same right-handed helicity as seen in plot (d). The upper and lower parts of the figure-8 spectrum are connected with each other by the complex arks extending via the boundary of the Brillouin zone at $p = \pm \pi$. Region III is sandwiched in between regions II and IV with the couplings satisfying $4 t_A t_B > h > - (t_A - t_B)^2$ and $t_B < - t_A$.
\item[$\triangleright$] \underline{Region V} (shaded by the green color) possesses the counterclockwise skewed 8-type spectrum that hosts two fermions with the same left-handed helicity, an example is shown in plot (g).  The region is determined by the relations $0> h > - (t_A - t_B)^2$ and $t_B > - t_A$.

\item[$\blacktriangleright$] The energy level $E = 0$ at $p=0$ is always double degenerate and the spectrum of the model always possesses a Dirac cone at the origin $p=0$. The Dirac cone corresponds to the real-valued dispersions in all regions except for region IV where the Dirac cone is entirely located in the imaginary space.
\end{itemize}

The skewed-8 spectrum with two Weyl excitations of the same helicity are realized in the non-Hermitian regions III and V.

An unwanted property of the model is the presence of the complex branches in the energy spectrum which implies that the system is not unitary and the total particle number is not a conserved quantity. This situation is quite typical in open systems described by non-Hermitian Hamiltonians~\cite{Bender:2007nj}.

If the whole energy spectrum were real then, topologically, the two states of the same helicity at the central Dirac cone would be complemented with other states of opposite helicity at another cone zone so that the net helicity of the lattice states in the compact Brillouin would be zero. However, in our model example the energy curves go into the imaginary dimension that allows then to avoid crossing and form another Dirac cone. Consequently, no compensating helicity states with real energy appear and the net helicity of the lattice fermion states is zero. 

On the other hand, the model becomes non-unitary because of the appearance of the complex energy branches. Alternatively, one also can remark that two  fermionic excitations of the same handedness with real-valued dispersions at $p=0$ are ``compensated'' by two fermionic modes of the other handedness with complex energy dispersions at $p = \pm \pi$ [cf. Figs.~\ref{fig:diagram}(d) and (g)]. The compensation modes are an unstable mode with Im$(\omega) >0$ and a decaying dissipative mode with Im$(\omega) <0$ which come always in pairs. 

Concluding our survey of the parameter space we notice that that the most interesting physics is realized in the non-Hermitian regions III and V in which the energy spectrum contains the closed real-valued Dirac cones which are connected by the complex energy arcs.

\subsection{Robustness of the single Dirac cone against local disorder}

We are going to demonstrate that the closed 8-shaped Dirac cone is essentially robust feature of our model. First of all, we notice that the very existence of the Dirac point in the $\PT$--invariant Hamiltonian~\eq{eq:H:chain2:explicit} does not depend on global variations of any of the parameters of the model~\eq{eq:H:chain2:explicit}. Indeed, the changes of parameters do not shift the position of the Dirac point in the Brillouin zone which always appears at the point $(p,E)=(0,0)$. The above statement can be supplemented by analytical calculations that demonstrate the double-degeneracy of eigenvalues at $p=0$. This fact can also be seen from the examples of the energy spectra in Fig.~\ref{fig:diagram} across the whole parameter space: the Weyl branches of the Dirac cones come always in pairs and they are never gapped. The cones are real-valued everywhere except for region IV in which the Dirac cone is entirely located in the imaginary space.  

We have also checked the robustness of the Dirac spectrum against the local disorder. To this end we have disordered the Hamiltonian~\eq{eq:H:chain2:explicit} at each site,
\beqn
& & H_{\mathrm{disord}} = - i \sum_{l=1}^N \Bigl[ g_{l,1} \bigl(  b^\dagger_l a_l - b^\dagger_{l-1} a_l \bigr) + g_{l,2} \bigl( a^\dagger_{l+1} b_l - a^\dagger_l b_l\bigr) \nonumber \\
&&  + t_{l,A} (a_{l+1}^\dagger a_l - a_{l-1}^\dagger a_l) + t_{l,B} (b_{l+1}^\dagger b_l - b_{l-1}^\dagger b_l) \Bigr], \quad \quad
\label{eq:H:chain2:disordered}
\eeqn
where the disorder in the site-dependent couplings
\beqn
\begin{array}{rcl}
t_{l,A} & = & t_A (1 + \delta_{l,A}), 
\qquad
g_{l,1} = + \sqrt{|h|} (1 + \delta_{l,1}), \\[2mm]
t_{l,B} & = & t_B (1 + \delta_{l,B}), 
\qquad
g_{l,2} = - \sqrt{|h|} (1 + \delta_{l,2})
\end{array}
\qquad
\label{eq:disorder:couplings}
\eeqn
is simulated by $4N$ independent factors $\delta_{l,i}$ which fluctuate randomly and independently at each site $l=1,\dots,N$ at each type of the coupling $i = A, B, 1, 2$. The fluctuations are constrained to the range
\beqn
- d \leqslant \delta_{l,i} \leqslant d \,, \quad i = A, B, 1, 2, \quad l=1,\dots,N, \qquad
\label{eq:disorder:delta}
\eeqn 
where the nonnegative parameter $d \geqslant 0$ determines the degree of the local disorder.

\begin{figure}[thb]
\begin{center}
\includegraphics[scale=0.3,clip=true]{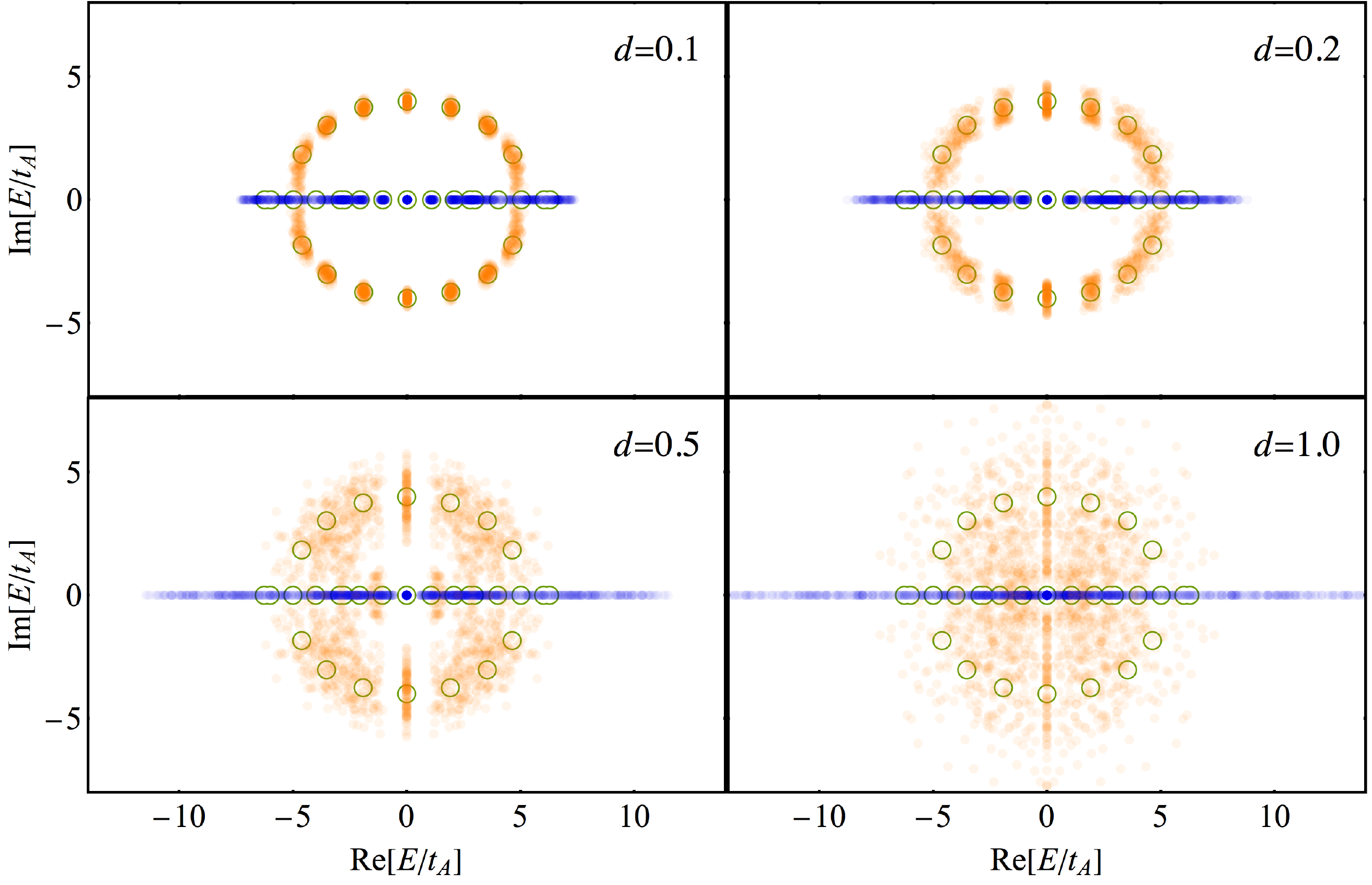}
\end{center}
\caption{The eigenvalues of the locally disordered Hamiltonian~\eq{eq:H:chain2:disordered} with the degrees of disorder $d=0.1, 0.2, 0.5$ and~$1$ for the chain of the length $2N$ with $N = 16$ for the non-Hermitian central parameters $t_B = 4 t_A$ and $g_1 = - g_2 = 2 t_A$ corresponding to the counterclockwise ``figure-8'' cone in plot (g) of Fig.~\ref{fig:diagram} (region V of the parameter space). The complex- and real-valued energies are shown in the orange and blue colors, respectively. The green open circles correspond to the undisordered Hamiltonian with $d=0$.}
\label{fig:disorder}
\end{figure}

In Fig.~\ref{fig:disorder} we show the distribution of the energy eigenvalues of the disordered Hamiltonian~\eq{eq:H:chain2:disordered} for various degrees of disorder $d$ at the periodic chain lattice of the length $32$. The parameters fluctuate randomly around the central values $t_B = 4 t_A$ and $g_1 = - g_2 = 2 t_A$ which correspond to the non-Hermitian region V of the parameter space, Fig.~\ref{fig:diagram} which exhibits the skewed figure-8 Dirac cone [a relevant example is shown in plot (g) of Fig.~\ref{fig:diagram}]. 

The zero-energy eigenvalue $E = 0$ at $p=0$ is always present at all disordered configurations of the couplings so that the presence of the Dirac cone is not affected by the disorder. While the disordered couplings do not open the gap they may turn the Dirac cone from the real-valued plane to the complex values plane. 

One can see that the moderate disorder with $d=0.1$ and $d=0.2$ perturbs only slightly both the central real-valued Dirac cone (the blue dots) and the connecting complex-valued arcs (the orange dots) as compared to the corresponding undisordered values (the green open circles). In other words, the excitations with the same helicity are not affect by the moderate disorder at all.

As expected, at larger disorder with $d=0.5$ and $d=1$, the central Dirac cone turns in the complex plane so that the spectrum shifts from region V to the totally unstable region IV.

\section{Conclusions}

We demonstrated that one-dimensional $\PT$-invariant non-Hermitian tight-binding systems may contain fermionic excitations with a nonvanishing net handedness and with the real-valued energy spectrum. We provided an explicit example~\eq{eq:H:chain2:explicit} of a local $\PT$-invariant non-Hermitian Hamiltonian with the desired energy spectrum. In particular, we demonstrated that the Brillouin zone may contain only {\emph{one}} Dirac cone with two Weyl fermions of the same handedness. The real-valued branches of the spectrum form a closed Dirac cone which resembles visually a skewed figure-8 curve. 

An example of the closed Dirac cone with two chiral modes of the same handedness is visualized in Fig.~\ref{fig:spectrum:example}. The upper and lower parts of the real spectrum are connected by two complex energy arks that automatically exclude the appearance of a compensating Dirac cone with a real-valued energy dispersion. The complex energy arks appear in a region of the Brillouin zone where the $\PT$ symmetry of the Hamiltonian is broken spontaneously. 

The model has the rich phase diagram shown in Fig.~\ref{fig:diagram}. The single closed Dirac cones are realized in two pockets of the parameters space (regions III and V in Fig.~\ref{fig:diagram}) at the non-Hermitian part of the phase diagram. In other words, the Hamiltonian must be non-Hermitian in order for the figure-8 dispersion to appear. 

We have also shown that the presence of the two chiral modes of the same handedness is robust agains a local random disorder of the Hamiltonian couplings. A moderate disorder does not open the gap in the 8-shaped Dirac cone which stays therefore protected. 

\acknowledgments

The author is very grateful to Mar\'ia Vozmediano for communications, valuable comments and suggestions, and encouragement.

\end{document}